\documentclass{iaus}
\usepackage{graphicx}
\usepackage{epsfig}
\usepackage{natbib}

\title[Radiative MHD in Massive Star Formation and Accretion Disks]
{Radiative Magneto-Hydrodynamics in \\ Massive Star Formation and Accretion Disks}

\author[Rolf Kuiper \& Mario Flock]   
{Rolf Kuiper$^1$, Mario Flock$^1$ \and Hubert Klahr$^1$}

\affiliation{
	$^1$Max-Planck Institut f\"ur Astronomie, \\
	K\"onigstuhl 17, 69117 Heidelberg, Germany \\ 
	email: {\tt kuiper@mpia.de}
}

\pubyear{2009}
\volume{259}  
\pagerange{100--100}
\date{"Dec. 09, 2008"  and in revised form ??}
\jname{Cosmic Magnetic Fields: From Planets, to Stars and Galaxies} 
\editors{K.G. Strassmeier, A.G. Kosovichev \& J.E. Beckman, eds.}

\begin{document}

	\maketitle

	\begin{abstract}
		We briefly overview our newly developed radiation transport module for MHD simulations 
		and two actual applications. 
		The method combines the advantage of the speed of the Flux-Limited Diffusion approximation 
		and the high accuracy obtained in ray-tracing methods.
		
	\end{abstract}	

	\firstsection 
	\section{Radiative MHD}
		
		Aim of the development of the radiation transport module described here is 
		to achieve a fast method for approximative frequency-dependent radiation transport 
		for (magneto-) hydrodynamical simulations in 
		studies of the environment around a single object source.
		
		The radiation transport module provides a Flux-Limited Diffusion (hereafter FLD) solver in cartesian, 
		cylindrical and spherical coordinates with a potentially non-equidistant grid spacing in the first dimension, 
		e.g. logarithmic in the radial coordinate. 
		Additionally to the FLD approximation the module consists of a first order ray-tracing technique 
		along the first dimension to account for single source irradiation 
		(e.g. stellar heating of a proto-planetary disk or radiative feedback in massive star formation).
			
		The local radiation field is split into two parts, 
		originated from an externally irradiated (frequency-dependent) flux $\vec{F}(\nu)$ 
		and a diffuse radiation energy density $E_R$, 
		which are in equlibrium with the radiation from the dust grains:
		\begin{eqnarray}
			a T^4 &=& E_R + \frac{1}{c \mbox{ }\kappa_P(T)} \int \kappa(\nu) \left|\vec{F}(\nu)\right| d\nu,
		\end{eqnarray}
		where $a$ is the radiation constant, 
		$c$ is the speed of light, 
		$\kappa(\nu)$ and $\kappa_{P}(T)$ represent the frequency-dependent 
		and the Planck mean opacity for a given temperature $T$ respectively. 
		Gas and dust temperatures are assumed to be the same. 
		The irradiation is treated as an instantaneous source of additional energy, 
		calculated via a first order ray-tracing as a function of the optical depth $\tau_\nu(r)$ 
        and distance $r$ from the central star:
        \begin{eqnarray}
            \vec{F}(\nu,r) &=& \vec{F}_*(\nu)  \mbox{ } \left(\frac{R_*}{r}\right)^2 \mbox{ } e^{-\tau_\nu(r)}
        \end{eqnarray}
			
		\noindent
		The evolution of the radiation energy is described by a Flux-Limited Diffusion equation
		\begin{eqnarray}
			\partial_t E_R &=& - f_c \left( \vec{\nabla} \cdot \left( D \vec{\nabla} E_R + \int \vec{F}(\nu,r) d\nu \right) - Q^+ \right)
		\end{eqnarray}
		with 
		$f_c = (c_v \rho / 4 a T^3 + 1)^{-1}$, $D = \lambda c / \kappa_{R} \rho$ 
		and 
		$Q^+ =  - p \vec{\nabla} \cdot \vec{v} + $additional source terms from hydrodynamics.
		The flux-limiter $\lambda$ is chosen according to 
		\cite{Levermore_Pomraning_1981}, 
		$c_v$ is the speciÞc heat capacity, 
		$\rho$, $p$ and $\vec{v}$ the gas density, thermal pressure and dynamical velocity 
		and $\kappa_R$ specifies the Rosseland mean opacity. 
		Scattering is neglected.

	\section{Application I: Massive Star Formation}
		
		Due to the short Kelvin-Helmholtz contraction timescale of a massive star 
		(\cite{Shu_1987})
		the accretion process onto such a star is described by the interaction of the gravitationally 
		forced inflow of matter with the radiative force escaping from the newly born star. 
		The conservation of angular momentum leads to the formation of an accretion disk as well as polar cavities. 
		At present we're studying this collapse scenario 
		with respect to the effect of dimensionality (from 1 to 3D) 
		and different applied physics 
		(isothermal and adiabatic test runs, 
		realistic cooling 
		and frequency-averaged as well as frequency-dependent radiative feedback).

	\section{Application II: Accretion Disks}
\begin{figure}[ht]
			\begin{center}
				\hspace{-1.8cm}
				\begin{minipage}[ht]{5.5cm}			
					\psfig{figure=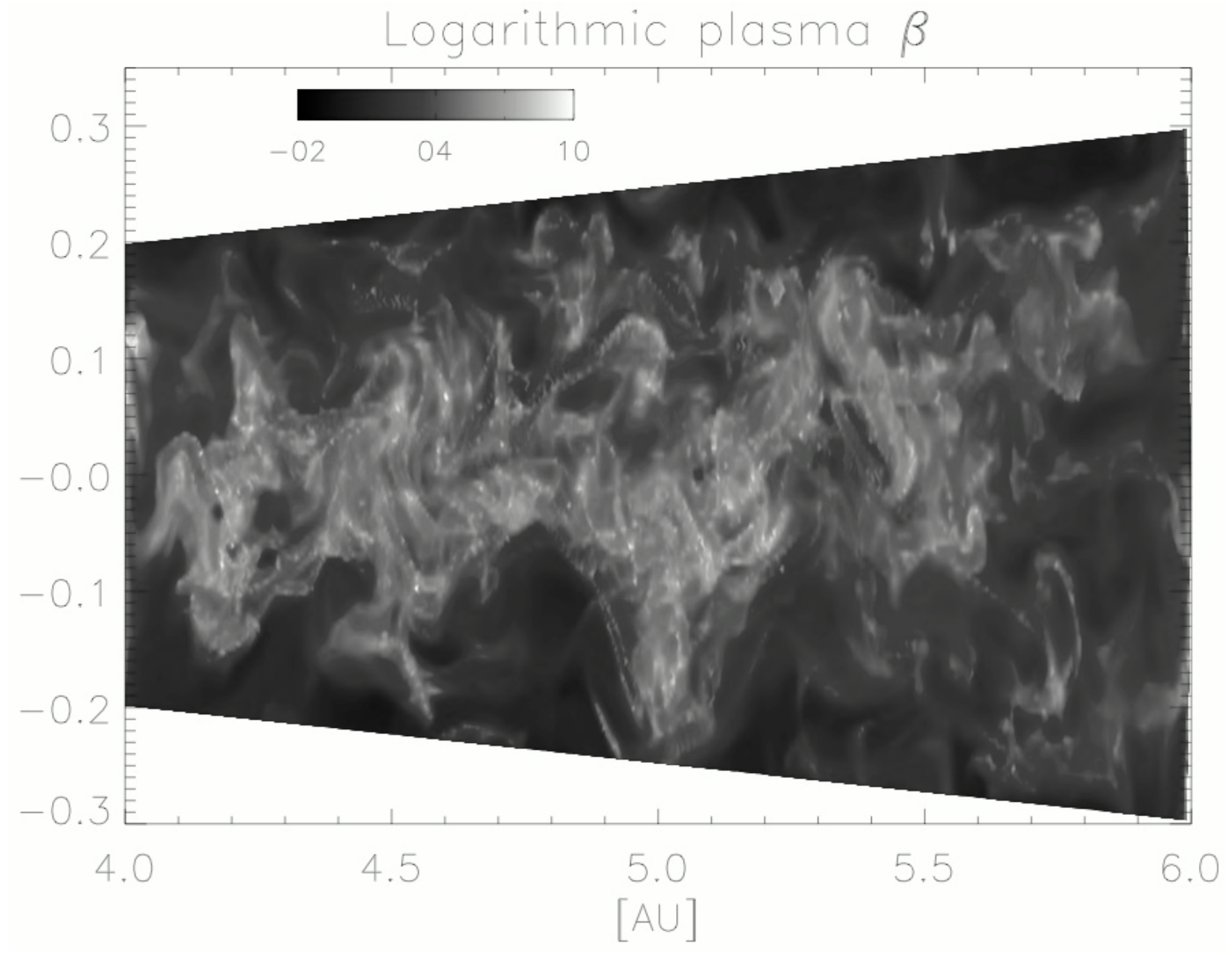,scale=0.25} 
				\end{minipage}
				\hspace{0.5cm}
				\begin{minipage}[hb]{5.5cm}			
					\psfig{figure=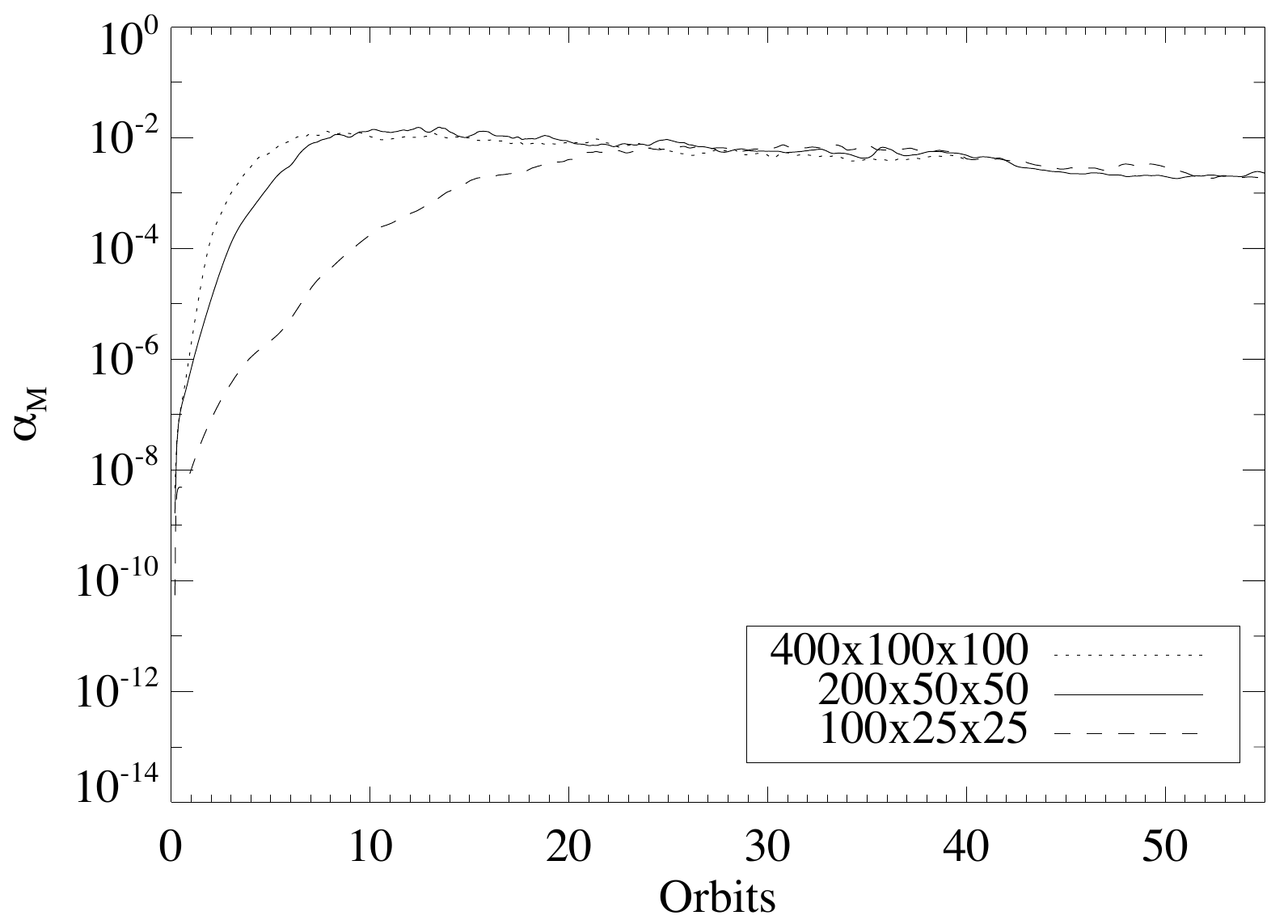,scale=0.30} 
				\end{minipage}
			\caption{
			     Left: Azimuthally averaged logarithmic plasma beta in the turbulent phase. 
			     \newline
			     Right: Evolution of Maxwell alpha at different resolution, 
			     converging against each other.
			} 
			\label{fig2}
			\end{center}
		\end{figure}

		We perform accretion disk simulations with a hydrodynamical stable stratified disk model 
		$\left(\frac{H}{R} = 0.05 \right)$ and an initial toroidal magnetic field. 
		The initial plasma beta $\beta$ is constant everywhere in the disk at a value of 25, 
		the runs by 
		\cite{Fromang_Nelson_2006} 
		(hereafter FN). 
		The convergence tests have a radial extension from 4 to 6 AU, $\pm$ 1 scale height 
		as vertical and $\pi/3$ as azimuthal extension. 
		We choose perodic boundaries in vertical and azimuthal direction. 
		Small velocity perturbations in radial and vertical direction about $10^{-4} c_0$ initiate the 
		non-linear MRI evolution. 
		With the used high-order MHD Riemann solver 
		(\cite{Mignone_2007}) 
		we get a converged alpha value of about $10^{-2}$ 
		presented in figure \ref{fig2}. 
        In opposite, FN got an increasing alpha value from low $\alpha=10^{-3}$ to high $\alpha=5*10^{-3}$ 
        (yet still not converged)
        with increasing resolution for a non-Riemann solver.
		In the future we will use the radiative module 
		to calculate proper resistive terms $\eta(T,\rho)$ 
		and to handle correctly the resulting heating 
		and cooling processes of MHD turbulence and the behaviour for the alpha stresses.

\end{document}